# The approach with the Data Protection and Privacy Relationships Model (DAPPREMO)[*]

Nicola FABIANO

Studio Legale Fabiano

Rome (Italy)

Affiliation: *International Institute of Informatics and Systemics (IIIS)* - USA

15/7/2021

**Abstract**

We describe the Data Protection and Privacy Relationships Model (DAPPREMO), which is based on the set theory, considering that both the data protection and privacy regulation and Ethics principles in those domains belong to a set. DAPPREMO is a new and innovative solution to adopt a model in any data protection and privacy activities. We theorise that DAPPREMO is an innovative approach to have a broad overview of all the objects related to a specific case or more cases from data protection and privacy perspective. We describe DAPPREMO as a solution for a multidisciplinary approach to address any data protection and privacy issue.

## 1. Introduction

Our confrontation with the protection of personal data and privacy is a constant note in our daily life. We cannot do without it (even when we face with exceptions that exclude the applicability of the specific legislation, just think for example of the hypothesis of state secrecy or something else). There is no aspect of our lives should not be assessed in terms of the impact on privacy and data protection. We saw this during the spring/summer 2020 health emergency that disrupted our lives and questioned our privacy and the protection of our personal data.

Their importance is unquestionable so much so that they are the subject of internal and European legislation which has the declared aim of ensuring the primacy of fundamental rights relating to the confidentiality and protection of individuals about the processing of personal data.

However, we should note that some contexts - particularly digital ones - seem to be governed by specific dynamics. Let us look, for example, at an Internet of Things (IoT) system: on closer inspection, it could be qualified only through the focus on the architecture, the technological solutions adopted and, of course, the devices used. In an IoT system, the technical component and, in particular, the specificity constituted by the communication between objects, would seem to be the only dominant and qualifying element. As we highlighted at the beginning of this chapter, we can describe IoT architecture as an "ecosystem" within which several components coexist, including the protection of personal data and privacy.

---

[*]This paper is an updated version of the chapter published in the book (2020) entitled "GDPR & Privacy. Awareness and opportunities. The approach with the Data Protection and Privacy Relationships Model (DAPPREMO)" [17].

In this IoT system (and here we would like to recall our arguments presented in other scientific fora briefly) is relevant is the protection of personal data (or privacy) topic because it becomes an essential part of the ecosystem. Data protection (or privacy) affects the entire ecosystem since it is precisely the latter aspect that is incredibly important. An IoT ecosystem without the data protection component does not perform all its functions correctly and is therefore imperfect.

The same consideration concerns any public or private context: the performance of ordinary activities - with or without the aid of technology - must be carried out in compliance with the regulations on the protection of personal data and privacy. One thinks, for example, of a public office that must carry out bureaucratic activities, or a private office that carries out sales activities, regardless of the use of IT tools. Indeed, the typical activities of each sector make up the core. Still, in both cases (public and private), relations with other contexts, such as the processing of personal data and confidentiality, can also be achieved. It is clear that even here, confidentiality and protection of personal data continue to play a fundamental role because the data controller must comply with the rules in force.

In essence, it emerges that confidentiality and the protection of personal data are of such importance that they significantly affect the context and are fundamental elements.

Given these premises, it is possible to reflect on the possibility of using the logical-mathematical analysis to qualify individual areas or domains. We use the term "domain" to refer to a specific sector, such as, for example, the core business of an activity, or the complex of processes of a Public Administration office, or even a single software development project.

It is also possible to make a leap and to identify a precise model that is useful for in-depth analysis and study, on the one hand, and for forecasting activities, on the other. It is necessary to proceed with mathematical logic, observing the reality consisting of activities for each sector and proposing a logical-mathematical reading. Each area corresponds to the whole (domain); objects are part of the entire (static or dynamic elements such as activities which, therefore, constitute processes).

Here is an example: a public administration office must carry out the activities that are specific to that sector, and each of them can be considered the object of a whole.

Let's make another example, similar: in a private company in the single business field, each activity can be considered an object that, together with the others (i.e., as a whole, all the actions), contribute to form a whole.

At this point, we can close the circle and superimpose this logical-mathematical observation considering the impact of the rules on the protection of personal data in each area, which from now on, we can classify as a whole.

The starting point is to ask whether the confidentiality and protection of individuals concerning the processing of personal data can also constitute an ecosystem, a whole. We believe that the answer is yes because they are both based on legal and ethical rules, which can form the objects of a system, of a whole. This set, however, is not an end in itself and cannot in itself generate activities. Still, its existence is necessary and functional to other sets (i.e. different contexts) with which it must relate. Indeed, a set of legal rules, although sector-specific, does not autonomously involve the performance of activities except through their concrete application. Regarding the protection of personal data, the provisions of the GDPR, therefore, autonomously do not involve the performance of activities. Still, in each case, the subjects, according to their respective roles, will be required to comply with this sector discipline.

Indeed, the protection of personal data does not merely consist of the respect and application of rules (this would be a minimal and limited view). In reality, data protection represents a real system, also defined, according to the mathematically oriented perspective, as a homogeneous set of objects (the single norms). Those objects could also coexist with others not present within the normative body of the sector (concerning to the protection of personal data, not expressly contained in a legal norm) of different nature (e.g., ethics), so to constitute a heterogeneous whole [see below].

## 2. Application of the set theory

In reality, there are almost daily interactions between systems, or domains, or themes (e.g., the Internet of Things "system" interacts with the data protection one, or the Public Administration "system" interacts with the data protection one and so on). In essence, schematically, it is possible to identify several sets, each belonging to a specific sector and composed of objects. These sets can express "one-to-one" or "one-to-many" relations between or among them, and also present possible sub-relations between the objects that contain them. Therefore, they are dynamic phenomena, not static, which involve continuous relationships and related effects.

It is essential to understand the mathematical explanation of relationships.

In mathematics, given two sets $A$ and $B$, where both $A$ and $B$ are not empty set, if we say that the set $A$ consists of objects that we state by $a, b, c, d, e, \ldots$ then we will denote $A = \{a, b, c, d, e, \ldots\}$, while if we say that the set $B$ may consist of totally different elements that we state by $x, y, s, t, u, v, \ldots$ then we will denote $B = \{x, y, s, t, u, v, \ldots\}$.

If we consider the following two sets:
$$A = \{a, b, c, d, e, \ldots\} \qquad \text{and} \qquad B = \{x, y, s, t, u, v, \ldots\}$$

any relation between $A$ and $B$ is usually denoted by writing notations such as $A\mathcal{R}B$, or $A\mathcal{E}B$, or even $A \sim B$ or with other symbols interposed between A and B. To define what exactly a relation between two sets is in Mathematics, however, we must consider the Cartesian product between two sets $A$ and $B$. This is, by definition, the new set indicated by the symbol $A \times B$ made by all the possible ordered pairs whose first co-ordinate is any element of $A$, and whose second co-ordinate is whichever element of $B$. For example, if $A = \{1, 2, 3, 4\}$ and $B = \{x, y\}$, then the Cartesian product $A \times B$ is the following set:

$$A \times B = \{(1, x), (1, y), (2, x), (2, y), (3, x), (3, y), (4, x), (4, y)\}$$

whose elements are all ordered pairs $(c_1, c_2)$, where $c_1$ is any element of $A$ (so $c_1 = 1$, or $c_1 = 2$, or $c_1 = 3$, or $c_1 = 4$) and $c_2$ is any element of $B$ (so $c_2 = x$, or $c_2 = y$).

Having acquired the notion of Cartesian product between two sets $A$ and $B$, we can define the notion of *relation* between $A$ and $B$. In Mathematics, a *relation* between $A$ and $B$ is defined to be any subset of the Cartesian product $A \times B$.

Therefore, if $A$ and $B$ are the two sets in the example just illustrated, an example of *relation* between $A$ and $B$ is given by the set
$$\mathcal{R} = \{(1, x), (1, y), (3, x)\}$$

In this relation, we have that the element $1 \in A$ is related to both $x$ and $y$, the latter being elements of $B$. The element $3 \in A$ is only related to $x$. Elements $2 \in A$ and $4 \in A$ are not related to anything (it is possible to assign relations in which one or more elements of the first set A are "orphans", i.e. are not related to anything). The relation
$$\mathcal{R} = \{(1,x), (1,y), (3,x)\}$$
can have a suggestive graphical representation if we use Venn diagrams and indicate the ordered pairs of elements in relation by arrows, as in the following figure.

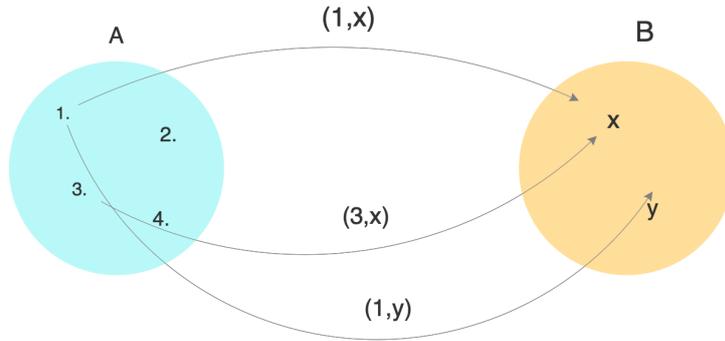

Given any relation $\mathcal{R}$ between two sets $A$ and $B$, the fact that the pair $(a, b)$ belongs to $\mathcal{R}$, i.e., which is the same, the fact the elements $a \in A$ and $b \in B$ are related by $\mathcal{R}$, is usually denoted by simply writing $a\mathcal{R}b$.

On the other hand, if we have more than two sets, for example a family of $n$ sets $(S_1, S_2, S_3, \ldots s_n)$, we will say that two of them, for example $S_i$ and $S_j$ are *connected* if there is a relation of any kind between them; we denote this configuration by $S_i \sim S_j$. In other terms, given a family of $n$ sets $(S_1, S_2, S_3, \ldots s_n)$ we have that:

$$S_i \sim S_j \iff \text{there exists a relation } \mathcal{R} \text{ such that } S_i \mathcal{R} S_j$$

Below is an image of the relation between several sets (one-to-many)

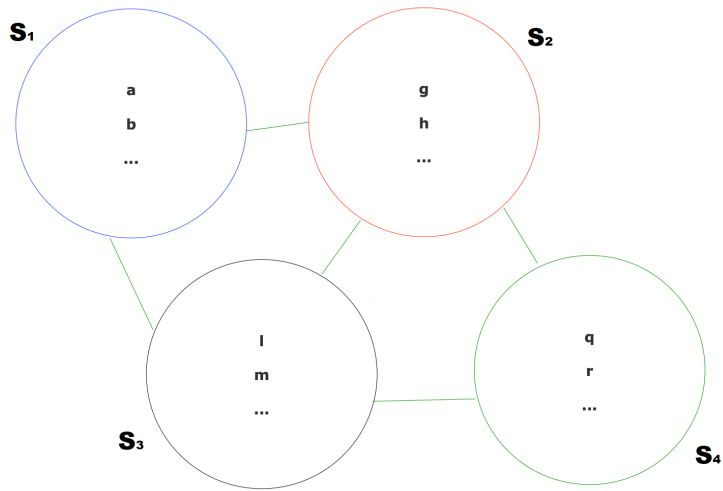

One set, say $S_1$, can be connected to several different sets at the same time. In this case, you have a one-to-many relationship.

That said, more specifically, we can describe the context related to data protection as a set of objects that are the legal rules of the sector in addition to other entities, even non-homogeneous (e.g. ethics), closely related to the context in question. The "data protection" set, therefore, contains the fundamental elements suitable to define the whole specific area. In essence, the protection of personal data has rules that are the norms for the sector, as well as any other field worthy of protection under the legal system. Therefore, the initial reference of the sector of personal data protection is constituted, precisely, by the field laws (for example, in Europe, the EU Regulation 2016/679 - GDPR [16]). However, even the hermeneutic survey of legal rules can reveal other essential aspects, such as ethics. These elements, legal norms and other entities contribute to constituting a real 'personal data protection' ecosystem that we can define as a whole. The 'data protection' set is identified by a characteristic property that unites all and only the elements. Therefore, objects of a set such as ethics and legal norms, have in common the characteristic property that is expressed as their applicative effect. Ethics, thus - which is not described in any data protection law - is an element of the whole because it has the same characteristic property as the other elements (the legal rules) in terms of their enforcement effect. The interaction with other sets does not end in a simple and, in any case, necessary verification of regulatory compliance. Still, it constitutes the initial impulse of an osmotic, exchange, and dynamic process that proceeds from the analysis of the specific context, ending up identifying all its attributes and, therefore, providing the necessary elements for the correct hermeneutic approach.

## 3. The relationships between objects and those between sub-assemblies

The above description constitutes a part - the initial one - which concerns the relations. Indeed, it is possible that, in addition to the connections between sets, there may also exist those between individual objects in a set, according to the following example scheme.

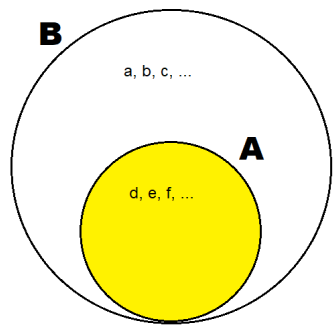

Considering this image and assuming that set A is the one related to privacy (or to the protection of personal data), it is evident not only what the regime of relations between sets maybe, but also what the significant impact of the subject matter covered by this volume with many other areas is. Indeed, any area generates a relationship with that relating to the protection of personal data and privacy. Identifying relationships by applying mathematics is a necessary added value to provide a broad and precise approach. The mentioned diagram shows the Eulero-Venn inclusion report which can be denoted by
$$A \subset B$$

Set $A$ relates to privacy (or protection of personal data) and the set $B$ describes a different area (or field or sector). It emerges that set $A$ is always present in the diverse relationships among sets since we cannot dismiss from the rules on the protection of personal data. It is an example of *subset*, where, as the number of objects in the set "privacy" in common with another set varies, the impact of the legal discipline and, therefore, the rules to be respected will vary accordingly. This graphic representation, expressed mathematically by the formula reproduced above, makes it possible to have a definite impact of the rules on the protection of personal data in each area, sector or area.

The inversion of the scenario just described shows how the set $A$ is the one related to the protection of personal data (or privacy) and broader then the smaller set $B$. It emerges that it appears feasible only in the hypothesis that exogenous elements or external objects (e.g. ethics) concerning the legal regulations on privacy are dominant and must also have in common with the smaller set or content. In particular, we said that the whole "protection of personal data" (or privacy), is mainly made up of legal regulations, is not an end in itself and, above all, does not enjoy autonomous operation, but is characterised by dynamic interaction with other sets.

By analysing the relationships, it is also possible that one may be created between sets and, in particular, and specifically between the one containing the legal rules on the protection of personal data and another domain, another area. It might come true a union of non-empty sets depending on the content (of the objects) that are shared by each set.

Turning back to the considerations of the beginning, we point out the fact that since the notation of inclusion $A \subset B$ implies that all elements of $A$ are also elements of $B$, when such an assumption is used in a context such as the one above, it implies that all legal and/or regulatory or operating/administrative rules of a specific sphere are also rules of privacy or ethics; in general, this may be too strong a claim and overly restrictive. Therefore, it might be helpful to express a similar concept but in a weaker, and thus broader, form, as follows. Let us denote by $P$ and $E$ the following sets:

$$P = \{\text{privacy rules}\} \quad \text{and} \quad E = \{\text{ethical principles}\}$$

Thus, setting $B = P \cup E$, it can be said that whatever the set $A$ of norms of a particular area of public life is, there is always some appropriate subset $A' \subset B$ which is related to $A$ according to the above sense of set relation. In formula:

$$\text{it exists } A' \subset B \text{ such that } A \mathcal{R} A'$$

Suppose, for example, $A$ indicates the set of all the rules of regulation and behaviour that - for example - regulate the functioning of a collegiate body. In that case, it is not necessarily the case that the element of $A$ are also data protection or privacy and Ethics rules, as the notation $A \subset B$ would suggest. Instead, it may be that the elements of $A$ are related to appropriate norms of data protection or privacy and Ethics. The latter eventually depend on the elements of $A$ and constitute the subset $A'$ of $B$ to which $A$ is related.

There is, nevertheless, yet another way by which it would be possibile to intepretate the whole framework. The following Figure is the Eulero-Venn diagram representing intersections of sets

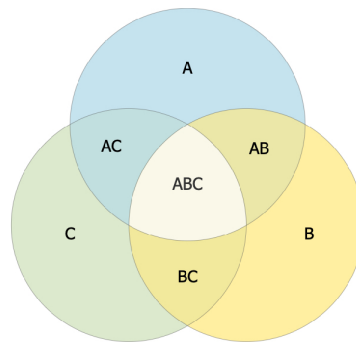

whose intersection relations are expressed mathematically as:

$$A \cap B \cap C, \quad A \cap C, \quad B \cap A, \quad C \cap B$$

Suppose that one of the three sets, for example, A, is the data protection (or privacy) set. The diagram represents an intersection between domains, one of which is the one related to the protection of personal data (or privacy), and it shows how it is possible to identify common areas constituted, precisely, by the intersections and therefore the relationships between sets and objects of each set or domain.

However, a clarification is necessary. What is proposed and described, even graphically, may seem represented on a two-dimensional plane. In reality, the complexity of the relationships must also be analysed on a three-dimensional or multidimensional plane, since in this way, it is possible to obtain a broader vision that is much closer to reality.

As it is evident, even in this brief exposition of the relations between sets, it emerges that one or more connections between different domains with a specific role of the one related to the protection of personal data (or privacy) can take place.

## 4. Description of a complex multidimensional model

The summary of the proposals described in the previous paragraphs leads to some reflections.

We said that the proposed model is based on set theory, where each area (e.g. privacy, personal data protection, Internet of Things, Public Administration sectors, private sectors, etc.) constitutes a set, and that is a domain. In everyday activities, there are always relationships between areas or domains. Indeed, the "protection of personal data" (which is a domain) has a relationship - for example - with that of a specific Publica Administration sector, or an IoT ecosystem, or a private sector. The model assumes greater complexity where the relationships between domains increase and may even tend to infinity.

The aim is to demonstrate that the application of this model, allowing a much broader view of the phenomena, allows a more precise evaluation of domains and individual relationships. As a result, there are, undoubtedly, beneficial effects for the entire system of analysis and especially concerning subjective profiles as will be illustrated later.

The activities carried out daily both for work, and personal needs are part (processes) of the system of the reality in which we live.

However, very often, activities and related processes are not correctly observed due to a short-sighted two-dimensional vision that is entirely reductive. The use of the most modern technological solutions has allowed us to see the so-called "augmented reality", i.e. multidimensional contexts. Therefore, if we observed phenomena not on a two-dimensional plane but a three-dimensional or multidimensional one, we would have the possibility to perceive with greater precision any component.

The request for a document to the Public Administration, for example, consists of two different processes: the application by a person, and the appropriate activities by the competent Office who have to handle it. Both the applicant and the Office carry out their actions on a single plane or level of observation. If one imagines using zoom and thus enlarging the field of view, the point of observation changes with the possibility of having an overall look from both the applicant's and the Office's side.

Well, if we use the same method for the analysis of scenarios relating to the relationship between the personal data protection (or privacy) domain and other domains, we would have a broader view with the opportunity to benefit from a completely different perception.

The complexity of the model proposed is precisely characterised by the innumerable and the unpredictable number of domains to be considered for relations, considering that in theory, the number of domains (i.e. sets) is potentially infinite.

The model appears very close, in its graphic representation, to that of a multidimensional distributed network system, where each point of intersection represents a set (a domain), and the union of the different points (which represents the sets) are the relationships. We should evaluate the representation not so much on a two-dimensional plane but in multidimensional terms as a link between sets. One of the essential elements is multidimensionality because we imagine different planes in space as if they were layers of a single system.

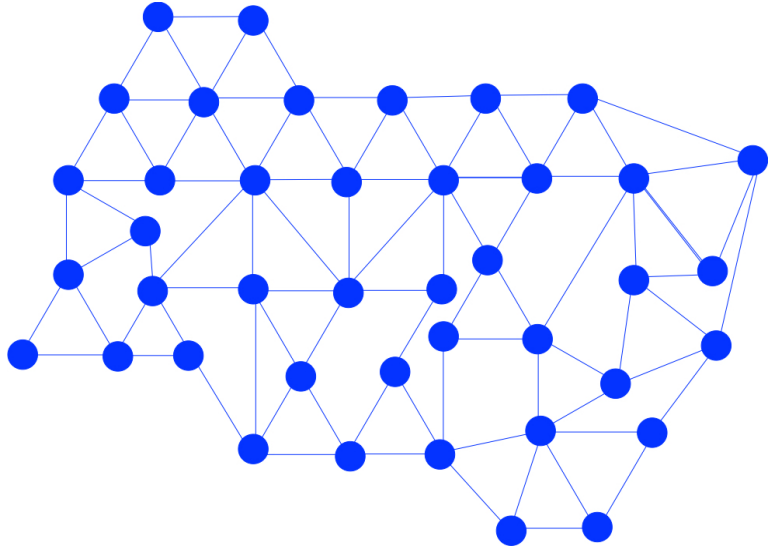

At a first, brief investigation, it emerges that the system of relations here presented could find some similarities with a complex structure, borrowed from advanced mathematics, known as the "fiber bundle set", which would seem to be able to account for the interactions and connections that occur both between individual elements and between sets of elements.

That is a first and innovative approach, in development, to a way of mathematical interpretation of the multidimensional inter-relational framework. Still, our model seems to have the merit of providing a unifying and abstract vision to a scenario of high complexity, such as the one on which we intend to focus our study investigation.

The complexity of the "**fiber bundle**" makes its description not simple. Still, we can illustrate it as a brush (see figure below) where the shaft represents, in our case, the data protection set, and the individual bristles constitute the relationships and connections between sets and objects of each other set[1].

We can also illustrate the "**fiber bundle**" with another image that is a book open at 360 degrees where the covers joined. That could give an idea of the complexity of the phenomenon.

The relationships are those existing between the single words present in the pages and represented by lines that intersect each other.

The proposed model, representing the links between sets, can be useful as a potent tool for a global and systemic approach.

The model, therefore, is based on the links between sets, where the objects belonging to the data protection (or privacy) domain are made up of the legislation in force and other external (exogenous) contexts to the legal discipline neither typical nor predetermined. The data protection set is related to the others, providing from time to time useful attributes for the qualification of the single scenario and therefore, the most suitable tools. It is not possible to generalise and predetermine the quality and quantity of attributes useful to the individual case, to the single scenario. These are sure enough dynamic and variable contexts that as much can differ from each other as, on the contrary, each one can achieve a standard in terms of approach and operation.

---

[1]This representation can be found on Wikipedia under "Fiber bundle".

The complexity of the model, at times, can make the dynamic aspect escape and lose sight of it, leaving space to focus only on the static part of the core that corresponds to the regulations in force. The regulatory discipline is not and cannot be an end in itself. Still, we must evaluate it as a contributing element to the analysis of a complex and dynamic context.

The attributes of the ecosystem data protection are numerous and, some precise and identified, and others indefinite and indeterminable. Among the specific fundamental objects, there is undoubtedly ethics, which is a crucial and essential element for the analysis of every single scenario. Ethics can seem an exogenous factor, extraneous to the set of rules governing the matter of the protection of personal data. In reality, this is not the case because the reference to ethics emerges from the principles contained in the whole body of rules. Thus, ethics is very close to the effects of the other elements of the whole (the legal norms) in terms of the same characteristic property that unites them.

The proposed model, which is called DAPPREMO, acronym of Data Protection Relationships Model, can be expressed mathematically through the concept of equivalence relationships.

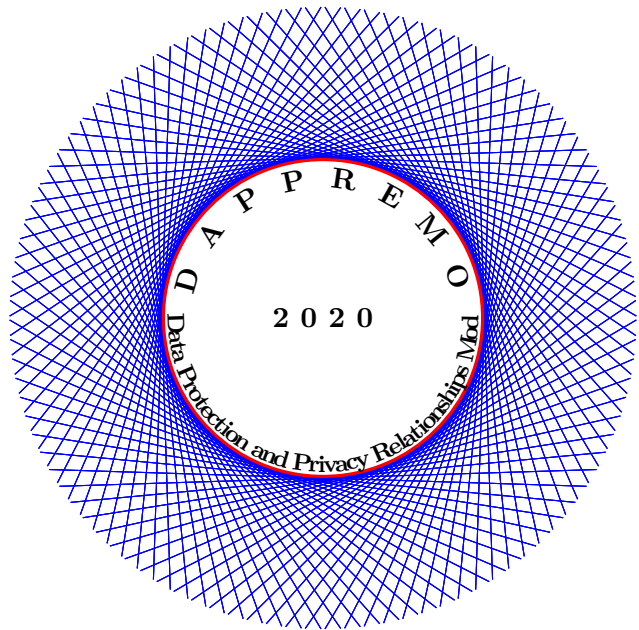

Analysing the results of the links and the relationships, the roles of the subjects and the activities on personal data, it is clear that many consequences arise from this model. Here we wanted to present the DAPPREMO model, although with the clarification that it is still the subject of further study and any updates will be contained in future publications.

## 5. The model and the role of the subjects

The DAPPREMO model, as described, allows having an extensive view of the relations between domains with the definite possibility of a detailed analysis and possible appropriate and specific solutions or interventions, much more precise than they commonly are.

We think that knowledge of the issues to be addressed and their full awareness are the basis of a correct approach to any scenario. In essence, the proposed model undoubtedly is a more incisive approach.

Being the case, what could be the effects of using this model for individual persons (let's call them "agents") involved in the data protection domain?

The first significant effect is on the supervisory authorities, since using the model-based approach described gives a comprehensive view, not only of the individual phenomenon but also of those related to it. Imagine that a supervisory authority has to face a preliminary investigation on a specific issue which, at first glance, is independent but is indeed strictly related to other domains.

An investigation, even if only exploratory, for example on the effects of an app, through the model-based approach, allows the supervisory authority to have a much broader view of any existing relationship (software development domain, Internet of Things, data transfer, etc.). If, on the other hand, the supervisory authority has to decide on a complaint, it has the possibility - through the proposed model-based approach - to have a comprehensive overview of each existing relationship with the case under examination.

In essence, it is a tool that offers the opportunity of an approach consistent with the "new" privacy, which must consider not only the regulatory part but also any kind of context that can be related to the phenomenon - even purely bureaucratic - under examination.

About the data controller and data processor, then, the use of the model will offer undoubted advantages in terms of analysis of the links of its core business with any other domain with which there is a connection.

In this way, the complex activity of adaptation and compliance with the legal rules on the "protection of natural persons with regard to the processing of personal data" will be facilitated and, above all, much more precise, knowing well in the analysis phase which types of principles to implement and the consequent interventions to be carried out.

It will also benefit the person concerned by using the model in terms of a different approach than merely consulting and applying legal rules. The knowledge of existing relationships regarding one's personal data allows one to exercise one's rights more consciously and correctly.

## Acknowledgement

A special thank you to professor Angelo Caldarella.